\documentclass[twocolumn,prc,floatfix,showpacs,preprintnumbers,nofootinbib,%
superscriptaddress]{revtex4}

\usepackage{graphicx} 
\usepackage{amsmath,amssymb}
\usepackage{slashed}
\usepackage{xcolor}
\definecolor{lcolor}{rgb}{0.5,0,0}
\definecolor{citcolor}{rgb}{0,0.3,0.0}
\usepackage[breaklinks,colorlinks,urlcolor=blue,citecolor=citcolor,linkcolor=lcolor]{hyperref}

\def\P{{\boldsymbol P}}
\def\a{{\boldsymbol a}}
\def\k{{\boldsymbol k}}
\def\l{{\boldsymbol l}}
\def\p{{\boldsymbol p}}
\def\q{{\boldsymbol q}}

\def \picwidth {0.48\textwidth}

\newcommand{\der}{\mathrm{d}}
\newcommand{\xt}{{{\boldsymbol x}_\perp}}
\newcommand{\yt}{{{\boldsymbol y}_\perp}}
\newcommand{\bt}{{{\boldsymbol b}_\perp}}
\newcommand{\rt}{{{\boldsymbol r}_\perp}}
\newcommand{\kt}{{\k_\perp}}
\newcommand{\lt}{{\l_\perp}}
\newcommand{\pt}{{\p_\perp}}
\newcommand{\qt}{{\q_\perp}}
\newcommand{\at}{{\a_\perp}}
\newcommand{\Pt}{{\P_\perp}}

\newcommand{\ptt}{P_\perp} 
\newcommand{\ktt}{k_T} 

\newcommand{\ud}{\, \mathrm{d}}

\newcommand{\tr}{\, \mathrm{Tr} \, }

\newcommand{\nc}{{N_\mathrm{c}}}

\newcommand{\da}{d_\mathrm{A}}
\newcommand{\nr}[1]{(\ref{#1})}

\newcommand{\qs}{Q_\mathrm{s}}

\newcommand{\qso}{Q_\mathrm{s0}}

\newcommand{\lqcd}{\Lambda_{\mathrm{QCD}}}
\newcommand{\as}{\alpha_{\mathrm{s}}}

\newcommand{\fig}{Fig.~}

\newcommand{\eq}{Eq.~}

\newcommand{\Jpsi}{{J/\psi}}

\begin{document}

\title{Forward $J/\psi$ production in proton-nucleus collisions at high energy}
\author{B. Duclou\'e}
\affiliation{
Department of Physics, University of Jyv\"askyl\"a %
 P.O. Box 35, 40014 University of Jyv\"askyl\"a, Finland
}
\affiliation{
Helsinki Institute of Physics, P.O. Box 64, 00014 University of Helsinki,
Finland
}

\author{T. Lappi}
\affiliation{
Department of Physics, University of Jyv\"askyl\"a %
 P.O. Box 35, 40014 University of Jyv\"askyl\"a, Finland
}
\affiliation{
Helsinki Institute of Physics, P.O. Box 64, 00014 University of Helsinki,
Finland
}

\author{H. M\"antysaari}
\affiliation{
Department of Physics, University of Jyv\"askyl\"a %
 P.O. Box 35, 40014 University of Jyv\"askyl\"a, Finland
}


\begin{abstract}
Inclusive production of $\Jpsi$ mesons, especially at forward rapidities, is an important probe of small-$x$ gluons in protons and nuclei. In this paper we re-evaluate the production cross sections in the Color Glass Condensate framework, where the process is described by a large $x$ gluon from the probe splitting into a quark pair and eikonally interacting with the target proton or nucleus. Using a standard collinear gluon distribution for the probe and an up to date dipole cross section fitted to HERA data to describe the target we achieve a rather good description of the cross section in proton-proton collisions, although with a rather large normalization uncertainty. More importantly, we show that  generalizing the dipole cross section to nuclei in the Glauber approach results in a nuclear suppression of $\Jpsi$ production that is much closer to the experimental data than claimed in previous literature.
\end{abstract}

 \maketitle

\section{Introduction}

Heavy quarks occupy an important place in QCD phenomenology. The mass of the charm quark is large enough that one should be able to describe its interactions using  weak coupling. On the other hand, it is light enough to be a dynamic probe that is sensitive to momentum scales that characterize high energy nuclear collisions, such as the deconfinement temperature or the gluon saturation scale. Because of the clean experimental signatures, vector meson production is a particularly important probe of heavy quark interactions. The ``melting'' of the $\Jpsi$ meson has traditionally been considered as one of the most important signals of the deconfinement transition in ultrarelativistic heavy ion collisions. Studying $\Jpsi$ production in proton-proton and proton-nucleus collisions at high energies can also provide  valuable insight into the physics of gluon saturation and strong color fields at small $x$, as well as into the energy loss of a high energy probe in nuclear matter.

We will in this paper use the Color Glass Condensate (CGC) framework together with a simple color evaporation model to calculate cross sections for $\Jpsi$ and $\Upsilon$ production at forward rapidities in proton-proton and proton-nucleus collisions at LHC energies. At forward rapidity the probe proton can be treated as a ``dilute'' collection of partons, and the CGC calculation proceeds according to a rather straightforward picture. The probe proton can be described by conventional collinear parton distributions in the so-called ``hybrid'' formalism, or by a  
corresponding ``$\ktt$-factorized'' approximation (which in fact strictly speaking
violates $\ktt$-factorization~\cite{Fujii:2005vj} since it involves several unintegrated gluon distributions). The physical picture in both cases is the same: a gluon from the probe can split into a quark-antiquark pair. As this system propagates through the target, it can pick up an eikonal phase described by an SU(3) color matrix, the ``Wilson line'' from the target. This leads to an expression for the $c\bar{c}$ pair production cross section (in either singlet or octet color states) in terms of Wilson line correlators describing the target. The same Wilson line correlators, the simplest one of them being the ``dipole cross section'', appear in calculations of e.g. total DIS cross sections~\cite{Albacete:2010sy,Rezaeian:2012ji,Rezaeian:2013tka},
single~\cite{Tribedy:2011aa,Albacete:2012xq,Rezaeian:2012ye,Lappi:2013zma} and double inclusive~
\cite{Lappi:2012nh,Albacete:2010pg,Stasto:2012ru,JalilianMarian:2012bd} 
particle production in proton-proton and proton-nucleus collisions,
diffractive DIS~\cite{Kowalski:2006hc,Lappi:2013am} 
and initial state for the hydrodynamical modeling of a heavy ion 
collision~\cite{Lappi:2011ju,Schenke:2012wb,Gale:2012rq}, making the framework very broadly applicable to many different processes.

Several LHC experiments have performed detailed measurements of inclusive $\Jpsi$ production in a broad range in rapidity, particularly in proton-proton~\cite{Khachatryan:2010yr,Aaij:2011jh,Aad:2011sp,Chatrchyan:2011kc,Abelev:2014qha} but also in proton-nucleus~\cite{Abelev:2013yxa,Aaij:2013zxa} collisions.
The results on the ``cold nuclear matter suppression'', i.e. the suppression of the cross section in proton-nucleus relative to proton-proton collisions, have then been compared to several theoretical predictions. In particular, the nuclear suppression predicted by the CGC calculation in~\cite{Fujii:2013gxa} has been much stronger than observed in the data. The main purpose of this paper is to explore to what extent this disagreement stems from the framework itself, and how much is due to approximations concerning nuclear geometry that are necessary for relating the CGC description of the proton to that of the nucleus. Indeed, in the case of single inclusive light hadron production~\cite{Lappi:2013zma}, the disagreement with early CGC nuclear suppression calculations~\cite{Albacete:2010bs} and LHC data turned out to be mostly due to the nuclear geometry effects.
We will in this paper calculate  cross sections for inclusive (prompt) forward $\Jpsi$ production in proton-proton and proton-nucleus 
collisions using the collinear ``hybrid'' framework also used in Ref.~\cite{Fujii:2013gxa}. 

We use here the dipole cross sections obtained in Ref.~\cite{Lappi:2013zma} 
by performing a fit to HERA data using the running coupling BK equation.
These have several advantages over the ones used in previous works.
Firstly the dipole cross section in a nucleus is related to the one in a 
proton taking into account the fluctuating positions of the nucleons in 
the nucleus via a Glauber model, without any additional parameters beside 
the standard Woods-Saxon nuclear density. This leads, unlike many previous
 CGC calculations,  to nuclear modification ratios that explicitly approach 
one in the high transverse momentum limit, at all collision energies. 
Physically this should be expected from the validity of collinear 
factorized pQCD in this regime. 
Secondly, the dipole cross section from 
Ref.~\cite{Lappi:2013zma}
corresponds to an explicitly positive unintegrated gluon distribution when Fourier-transformed to momentum space. Thirdly, our calculation, like that of \cite{Lappi:2013zma}, takes consistently into account also the proton transverse area from the fit 
to HERA data. This completely fixes the normalization of the LO calculation at the quark level without any additional parameters to describe the proton size\footnote{Note, however, that the normalization is expected to be quite different at NLO, and that for $\Jpsi$ in the color evaporation model the normalization is given by an additional phenomenological parameter.}.
The use of also the proton transverse area from the DIS fits and the
way of extending the dipole cross section
from a proton to a nucleus are quantitatively the most important differences 
between our work and that of~\cite{Fujii:2013gxa}.

This paper is organized as follows. We will first briefly describe the formalism used in the calculation in Sec.~\ref{sec:formulae} and the used dipole correlators in Sec.~\ref{sec:dipole}. We will then present our results for cross sections in proton-proton and proton-nucleus collisions in Sec.~\ref{sec:results} and finish with a discussion and outlook for future work in this direction in Sec.~\ref{sec:disc}.

\section{Formalism}
\label{sec:formulae}

We will here use the simple color evaporation model (CEM), where a fixed fraction of all $c\bar{c}$ pairs produced below the $D$-meson threshold are assumed to become $\Jpsi$ mesons. The quark pair can be produced in either the singlet or octet representation, the additional color is assumed to be ``evaporated'' away in the form of a soft gluon.
In the CEM model, the differential cross section for $\Jpsi$ production therefore reads
\begin{align} 
\frac{\ud\sigma_{\Jpsi}}{\ud^2\Pt\ud Y}
=
F_{\Jpsi} \; \int_{4m_c^2}^{4M_D^2} \ud M^2
\frac{\ud\sigma_{c\bar c}}
{\ud^2\Pt \ud Y \ud M^2}
\, ,
\label{eq:dsigmajpsi}
\end{align}
where $\Pt$ and $Y$ are the transverse momentum and rapidity of the produced $\Jpsi$, $\frac{\ud\sigma_{c\bar c}}{\ud^2\Pt \ud Y \ud M^2}$ is the cross section for the production of a $c\bar c$ pair with transverse momentum $\Pt$, invariant mass $M$ and rapidity $Y$, $m_c$ is the charm quark mass and $m_D=1.864$ GeV is the $D$ meson mass. Here $F_{\Jpsi}$ is a nonperturbative quantity representing the probability that a $c\bar c$ pair has to form a $\Jpsi$. In the literature
\cite{Mangano:2004wq} this constant is taken to have different values depending on the $c$-quark mass, the parton distribution and the factorization scale. We will here use a constant value $F_{\Jpsi}=0.06$. This is somewhat larger than the typical values in perturbative NLO calculations, but it should be kept in mind that our calculation is at the LO level and is therefore likely to underestimate the normalization of the $c\bar{c}$ cross section.
The same model can be used to compute the cross section for $\Upsilon$ 
production by simply replacing $m_c$ with the bottom quark mass, $M_D$ with
the $B$ meson mass 
$M_B=5.280$~GeV and $F_{\Jpsi}$ with $F_{\Upsilon}$ that we take equal to 0.04 here.
One should note that since the $b$-quark mass is already significantly 
higher than the saturation scale, bottom quark production 
is not in the natural regime of 
validity of the CGC framework. We will in this paper however calculate also 
$\Upsilon$ cross sections for comparison.

The formalism for analyzing gluon and quark pair production in the dilute-dense limit of the CGC formalism has been worked out in great detail already some time ago~\cite{Blaizot:2004wu,Blaizot:2004wv} (see also \cite{Kharzeev:2012py})
and applied in many calculations such as~\cite{Fujii:2006ab,Fujii:2005rm,Fujii:2013gxa,Fujii:2013yja}.
It can be obtained from the observation that an incoming gluon from the projectile can split into a quark-antiquark pair either before or after the interaction with the target. These partons are then assumed to eikonally interact with the target, picking up a Wilson line factor in either the adjoint or the fundamental representation, depending on the particle. We will here use the collinear approximation, where the incoming gluon is assumed to have zero transverse momentum.
Calculated in this way the cross section for the production of a $c\bar c$ pair with transverse momenta $\pt$ and $\qt$ and rapidities $y_p$ and $y_q$ reads, in the large-$\nc$ limit~\cite{Fujii:2013gxa},
\begin{multline}
\!\!\!\!\! \frac{\ud \sigma_{c\bar{c}}}{\ud^2\pt \ud^2\qt \ud y_p \ud y_q}
= 
\frac{\as^2 \nc}{8\pi^2 \da}
\frac{1}{(2\pi)^2}
\!\!\int\limits_{\kt}\!
\frac{\Xi_{\rm coll}(\pt + \qt,\kt)}{(\pt + \qt)^2}
\\ \times
\phi_{y_2=\ln{\frac{1}{x_2}}}^{q\bar{q},g}(\pt + \qt,\kt)
\;
x_1 G_p(x_1,Q^2) \; ,
\label{eq:dsigmaccbarcoll}
\end{multline}
where $\int_{\kt} \equiv \int \ud^2 \kt / (2\pi)^2$,
and $\da\equiv \nc^2-1$ is the dimension of the adjoint representation of
SU($\nc$). The longitudinal momentum fractions of the gluons coming from the projectile and the target, $x_1$ and $x_2$, are given by
\begin{equation}
	x_{1,2}=\frac{\sqrt{\Pt^2+M^2}}{\sqrt{s}}e^{\pm Y} \; .
\end{equation}
The ``hard matrix element'' 
$\Xi_{\rm coll}$ is the sum of 3 terms,
\begin{equation}
\Xi_{\rm coll}
=
\Xi_{\rm coll}^{q\bar{q},q\bar{q}}
+
\Xi_{\rm coll}^{q\bar{q},g}
+
\Xi_{\rm coll}^{g,g}\; ,
\end{equation}
corresponding to a quark pair scattering off the target ($q\bar{q},q\bar{q}$), 
the gluon scattering before splitting into a quark pair ($g,g$) and an interference between the two. The explicit expressions are
\begin{align}
\Xi_{\rm coll}^{q\bar{q},q\bar{q}}
=&
\frac{8p^+q^+}{(p^++q^+)^2(\at^2+m_c^2)^2} \nonumber\\
&\times\bigg[m_c^2+\frac{(p^+)^2+(q^+)^2}{(p^++q^+)^2}\at^2\bigg] , \nonumber\\
\Xi_{\rm coll}^{q\bar{q},g}
=&
-\frac{16}{(p+q)^2(\at^2+m_c^2)} \nonumber\\
&\quad\times\bigg[m_c^2+\frac{(p^+)^2+(q^+)^2}{(p^++q^+)^3}\at\cdot(p^+\qt-q^+\pt)\bigg] , \nonumber\\
\Xi_{\rm coll}^{g,g}
=&
\frac{8}{(p+q)^4}\!
\bigg[
(p+q)^2\!-\!\frac{2}{(p^++q^+)^2}(p^+\qt\!\!-q^+\pt)^2
\bigg] ,
\end{align}
with $\at\equiv \qt-\kt$.

The propagation of a quark-antiquark pair through the color field of the target
is described by the function
\begin{equation}\label{eq:defphi}
\phi_{_Y}^{q \bar{q},g}(\lt,\kt)=
\int\der^2 \bt \frac{N_c\lt^2}{4 \as} \; 
S_{_Y}(\kt) \;
S_{_Y}(\lt-\kt) \;,  
\end{equation}
where $\bt$ is the impact parameter.
All the information about the target is contained in the function $S_{_Y}(\kt)$, which  is the fundamental representation dipole correlator 
in the color field of the target:
\begin{equation}
 S_{_Y}(\kt) = \int \ud^2 \rt  
e^{i\kt \cdot \rt }
 S_{_Y}(\rt) \;,
\end{equation}
with
\begin{equation}
S_{_Y}(\xt-\yt) = \frac{1}{\nc }\left< \tr U^\dag(\xt)U(\yt)\right>,
\end{equation}
where $U(\xt)$ is a fundamental representation Wilson line in the target color field. Our calculation is a leading order one, so for consistency we should
use a leading order collinear gluon distribution to describe the probe. 
In this work we use the MSTW 2008~\cite{Martin:2009iq} LO parametrization
for $G_p(x_1,Q^2)$.

\begin{figure}[tb]
  \centering\includegraphics[width=\picwidth]{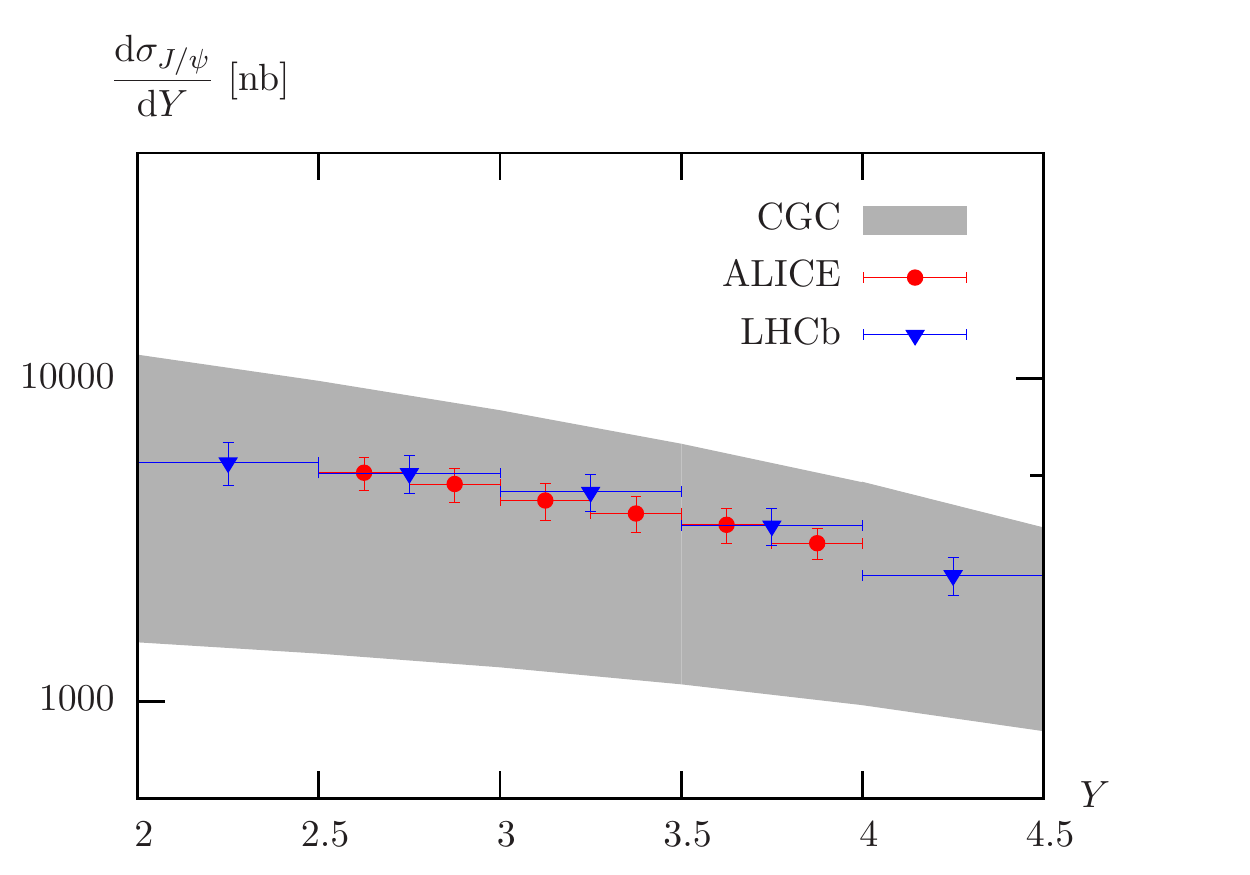}
  \caption{Differential $\Jpsi$ cross section as a function of $Y$ in pp collisions. Data from Refs.~\cite{Abelev:2014qha,Aaij:2011jh}.}
  \label{fig:dsigma_dY_pp_Jpsi}
\end{figure}

\begin{figure}[tb]
	\centering\includegraphics[width=\picwidth]{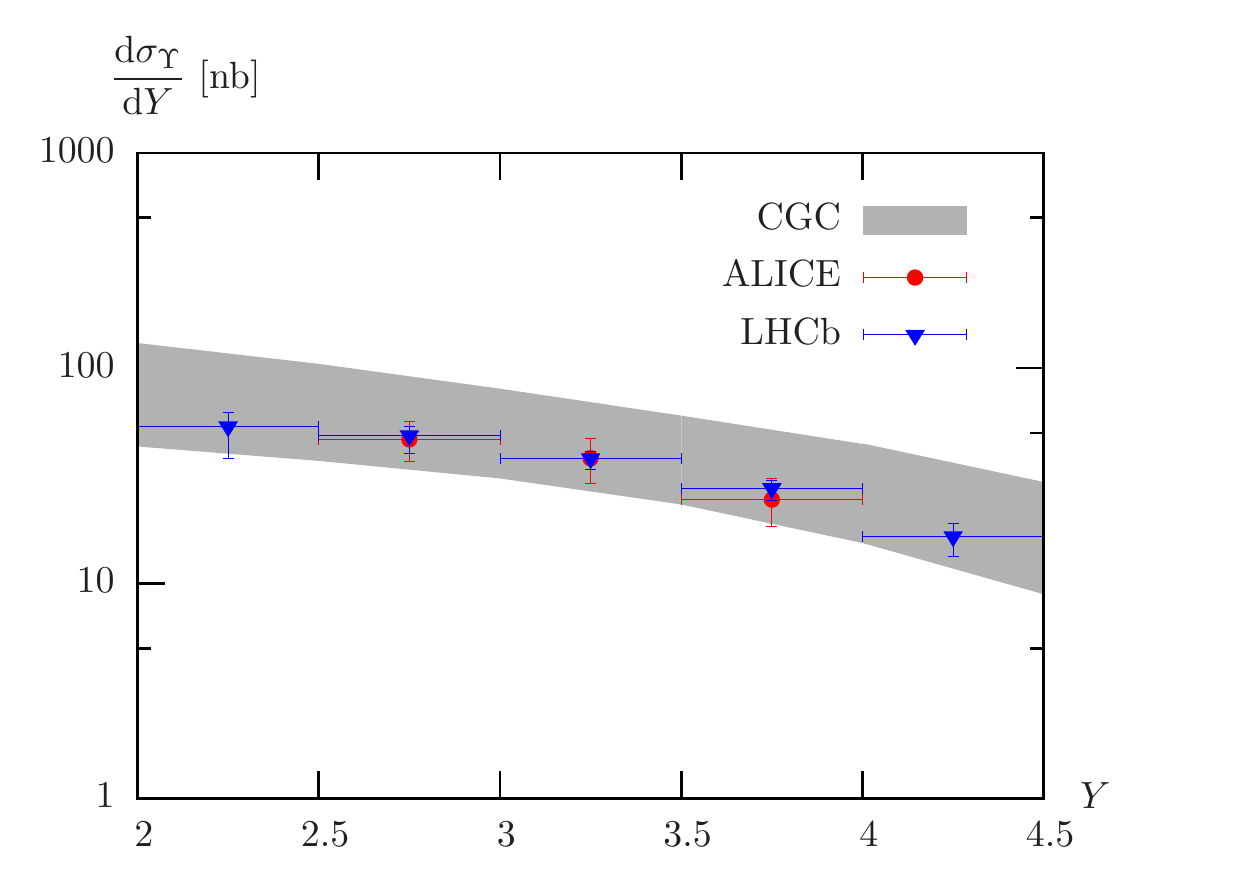}
	\caption{Differential $\Upsilon$ cross section as a function of $Y$ in pp collisions. Data from Refs.~\cite{Abelev:2014qha,LHCb:2012aa}.}
	\label{fig:dsigma_dY_pp_Upsilon}
\end{figure}

\begin{figure}[tb]
	\centering\includegraphics[width=\picwidth]{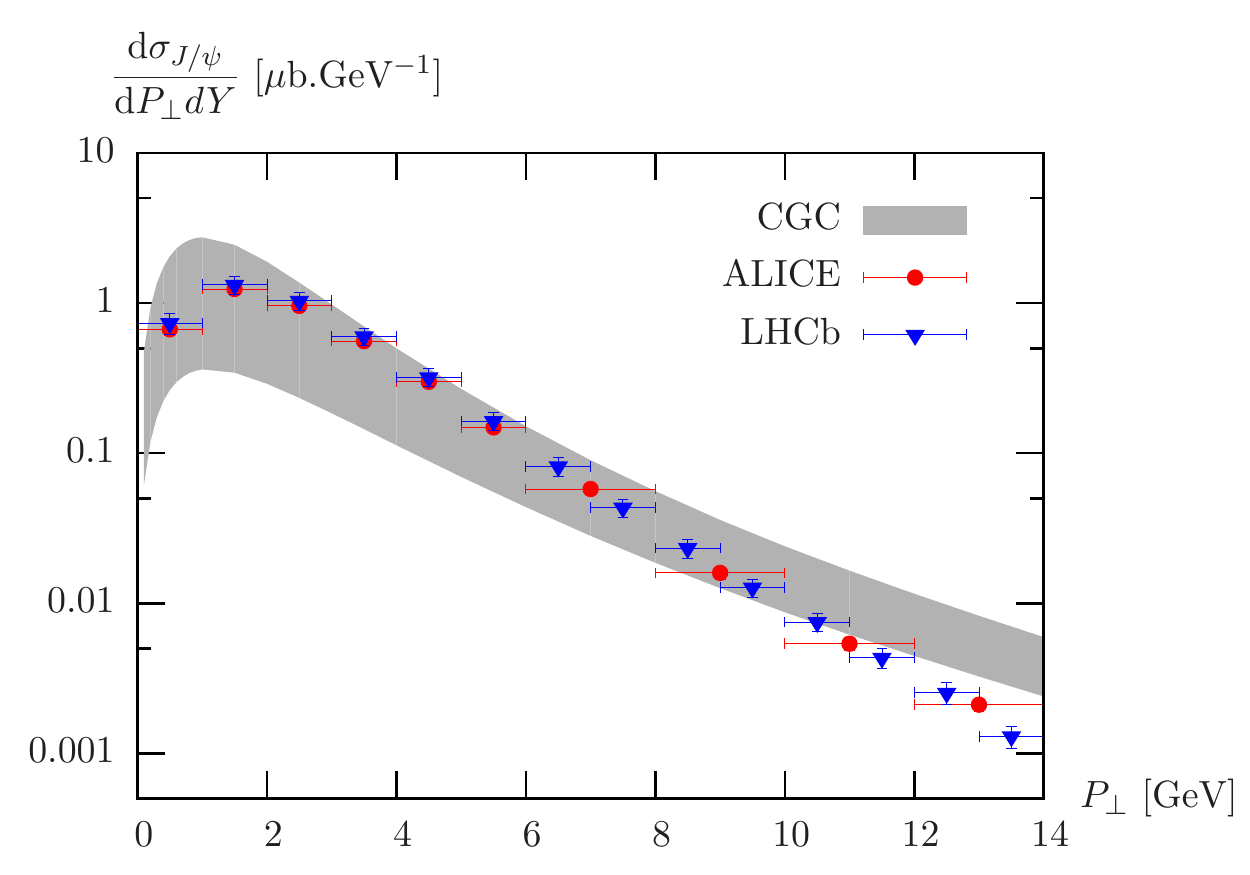}
	\caption{Differential $\Jpsi$ cross section as a function of $\ptt$ in pp collisions ($2.5<Y<4$). Data from Refs.~\cite{Abelev:2014qha,Aaij:2011jh}.}
	\label{fig:dsigma_dpT_dY_pp_Jpsi}
\end{figure}

\begin{figure}[tb]
	\centering\includegraphics[width=\picwidth]{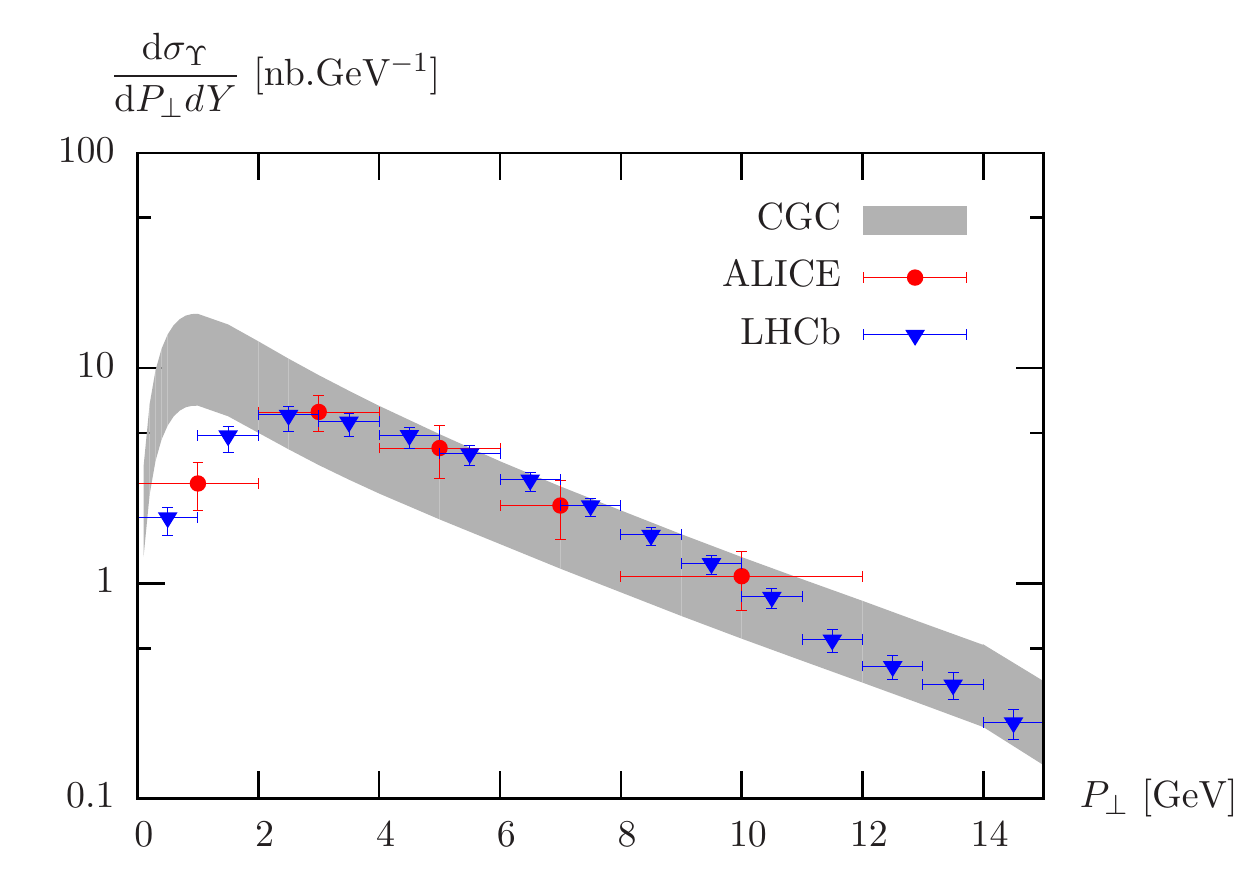}
	\caption{Differential $\Upsilon$ cross section as a function of $\ptt$ in pp collisions ($2.5<Y<4$). Data from Refs.~\cite{Abelev:2014qha,LHCb:2012aa}.}
	\label{fig:dsigma_dpT_dY_pp_Upsilon}
\end{figure}

\section{Dipole correlator}
\label{sec:dipole}
For the dipole correlator $S_{_Y}(\kt)$ we use the 
MV$^e$-parametrization from Ref.~\cite{Lappi:2013zma}. 
This is obtained by fitting the combined inclusive HERA DIS cross section data~\cite{Aaron:2009aa} in the region  $Q^2<50$ GeV$^2$ and $x<0.01$ using a dipole cross section, whose $x$-dependence is obtained with the running coupling BK equation~\cite{Balitsky:1995ub,Kovchegov:1999ua,Balitsky:2006wa}. 
The dipole amplitude at the initial rapidity $x_0=0.01$ is given by the parametrization
\begin{equation}\label{eq:icp}
S_{Y= \ln \frac{1}{x_0}}(\rt) = \exp \left[ -\frac{\rt^2 \qso^2}{4} \ln \left(\frac{1}{|\rt| \lqcd}\!+\!e_c \cdot e\right)\right].
\end{equation}
The impact parameter profile in the proton is assumed to factorize from the dimensionless dipole amplitude, and is parametrized by a constant 
\begin{equation}\label{eq:defsigma0}
\int\der^2 \bt \to \frac{\sigma_0}{2} \; ,
\end{equation}
so that the distribution needed for quark pair production is given by
\begin{equation}
\phi_{_p,_Y}^{q \bar{q},g}(\lt,\kt)
=
\frac{\sigma_0}{2} \; \frac{N_c\lt^2}{4 \as} \; 
S_{_Y}(\kt) \;
S_{_Y}(\lt-\kt).
\end{equation}
Using the expression 
\begin{equation}
        \as(r) = \frac{12\pi}{(33 - 2N_f) \log \left(\frac{4C^2}{r^2\lqcd^2} \right)}
\end{equation}
for the running coupling constant, the fit gives a very good description 
$\chi^2/\text{d.o.f} = 1.15$ of the data with the parameters
$\qso^2= 0.060$ GeV$^2$, $C^2= 7.2$, $e_c=18.9$  and $\sigma_0/2 = 16.36$ mb.
One  advantage of the parametrization~\cite{Lappi:2013zma} over the AAMQS~\cite{Albacete:2010sy} one used in~\cite{Fujii:2013gxa}
is that the Fourier transform of 
$S_{Y=\ln\frac{1}{x_0}}(\rt)$ is positive definite even at the initial rapidity,
enabling a more natural physical interpretation as an unintegrated gluon
distribution. We also used the MV and MV$^\gamma$ parametrizations from Ref.~\cite{Lappi:2013zma} for the initial condition of the BK equation and found that the effect of such a change is small compared with the other uncertainties involved in the calculation.

For the nucleus we use, again following~\cite{Lappi:2013zma}, a BK-evolved dipole correlator with the initial condition
\begin{multline}\label{eq:ica}
S^A_{Y=\ln \frac{1}{x_0}}(\rt,\bt) = \exp\bigg[ -A T_A(\bt) 
\frac{\sigma_0}{2} \frac{\rt^2 \qso^2}{4} 
\\ \times
\ln \left(\frac{1}{|\rt|\lqcd}+e_c \cdot e\right) \bigg] \; ,
\end{multline}
where
\begin{equation}
T_A(\bt)= \int dz \frac{n}{1+\exp \left[ \frac{\sqrt{\bt^2 + z^2}-R_A}{d} \right]} \; ,
\end{equation}
with $d=0.54\,\mathrm{fm}$ and $R_A=(1.12A^{1/3}-0.86A^{-1/3})\,\mathrm{fm}$, is the standard Woods-Saxon transverse thickness function of the nucleus,
normalized to unity. To calculate cross sections with nuclear targets 
we explicitly integrate over the impact parameter, treating the dilute edge of the nucleus (where the saturation scale of the nucleus falls below the proton saturation scale) as in Ref.~\cite{Lappi:2013zma}.  In practice this amounts to using the proton-proton result scaled such that the nuclear modification factor $R_{pA}$, defined in Eq.~\eqref{eq:defrpa}, is unity in the dilute edge of the nucleus. 
This ``Glauber model'' form can be derived by assuming that at the 
initial rapidity, a high energy 
probe scatters independently off the nucleons in the nucleus. The average 
over the fluctuating positions of the nucleons can then be performed 
analytically. At small $r$ the integral of \eq\nr{eq:ica} over $b$ explicitly approaches $A$ times the dipole-proton cross section~\nr{eq:icp}. Thus the form \nr{eq:ica} explicitly leads to nuclear modification
ratios $R_{pA}$ (see definition \eq\nr{eq:defrpa}) that, 
at all collision energies, 
approach unity at small distances, i.e. large momenta. 
We emphasize that besides the standard nuclear density $T_A(b)$ there 
are no additional parameters in passing from the proton, \eq\nr{eq:icp},
to the nucleus in \eq\nr{eq:ica}. Also the transverse area of the proton
$\sigma_0/2$ is consistently taken to be the same in 
proton DIS and in hadronic collisions, i.e. the 
identification~\nr{eq:defsigma0} is also used in the $pp\to \Jpsi + X$
cross section. 

In our calculations the nuclear 
modification only depends on parameters fit to proton data and on 
the standard nuclear thickness function. In particular, no independent
nuclear saturation scale with an unknown value is needed, in 
contrast to Ref.~\cite{Fujii:2013gxa}. The practical result of this is that 
the effective typical nuclear saturation scale from \eq\nr{eq:ica} is smaller
than the $A^{1/3}$-scaling from the proton assumed in 
Ref.~\cite{Fujii:2013gxa}. On the other hand, we treat separately the 
area occupied by the small-$x$ gluons in the proton, $\sigma_0/2$, 
and the total inelastic cross section, $\sigma_\textrm{inel}$.
 Thus here the $\Jpsi$
cross section in proton-proton collisions is proportional to 
 $\sigma_0/2$ which is smaller than the 
proton area assumed in Ref.~\cite{Fujii:2013gxa}. 
The net effect of the smaller nuclear 
$\qs$ and the smaller yield in proton-proton collisions is that 
one obtains a nuclear suppression ratio $R_{pA}$ 
that is closer to one around $\ptt \sim \qs$ 
than in Ref.~\cite{Fujii:2013gxa}, 
and simultaneously approaches unity at large $\ptt$.

\section{Results}
\label{sec:results}

In this section we will present the results of our calculation and show a 
comparison with LHC data from the ALICE and LHCb experiments in pp and pA 
collisions at center of mass energies of 7 and 5 TeV respectively. 
It should be noted that both for pp and pA collisions, LHCb gives separately
the cross section for prompt J/$\psi$ production and the contribution 
from b decays (which is not included in our calculation) while ALICE 
data is only for inclusive J/$\psi$ production. Therefore, for a
consistent comparison, we subtract the b decay fraction measured 
by LHCb from the ALICE data to extract the prompt contribution.

In the uncertainty band we include the variation of the charm
quark mass between 1.2 and 1.5 GeV (in the case of $\Jpsi$ production) or of the bottom
quark mass between 4.5 and 4.8 GeV (in the case of $\Upsilon$ production)
and of the factorization scale
between $2M_\perp$ and $M_\perp/2$ where 
$M_\perp=\sqrt{M^2+\Pt^2}$ and $M$ is the 
$q\bar q$ pair's invariant mass.
When we show quantities integrated over the transverse momentum of the produced meson, we integrate up to $\ptt=15$ GeV. Our results do not depend significantly on this choice since the cross section is quickly decreasing at large $\ptt$.

\subsection{Proton-proton collisions}

We show our results for the $\ptt$-integrated cross section 
$\frac{\ud \sigma_\Jpsi}{\ud Y}$ at a center of mass energy $\sqrt{s}=7$~TeV
in \fig\ref{fig:dsigma_dY_pp_Jpsi},  
compared with data from ALICE~\cite{Abelev:2014qha} and 
LHCb~\cite{Aaij:2011jh}. We see that the shape of the 
data is quite well described, but the normalization uncertainty is quite large.
Similar conclusions can be drawn for $\Upsilon$ production, as can be seen from \fig\ref{fig:dsigma_dY_pp_Upsilon} where we compare our results with ALICE~\cite{Abelev:2014qha} and LHCb~\cite{LHCb:2012aa} data, although we observe that the uncertainty band of our calculation is narrower than for $\Jpsi$ production.

In \fig\ref{fig:dsigma_dpT_dY_pp_Jpsi} we 
show the comparison of our results for the $\Jpsi$ $\ptt$ spectrum with ALICE and 
LHCb data. Again the agreement with the data is quite good 
except at very large transverse momenta where the CGC description is 
not expected to be very accurate.
For $\Upsilon$ production, as shown in \fig\ref{fig:dsigma_dpT_dY_pp_Upsilon}, the 
agreement is also quite good at higher $\ptt$, but this leading order framework
does not describe the increase in the  mean $\ptt$ of  the $\Upsilon$ compared
to the $\Jpsi$.

\begin{figure}[tb]
	\centering\includegraphics[width=\picwidth]{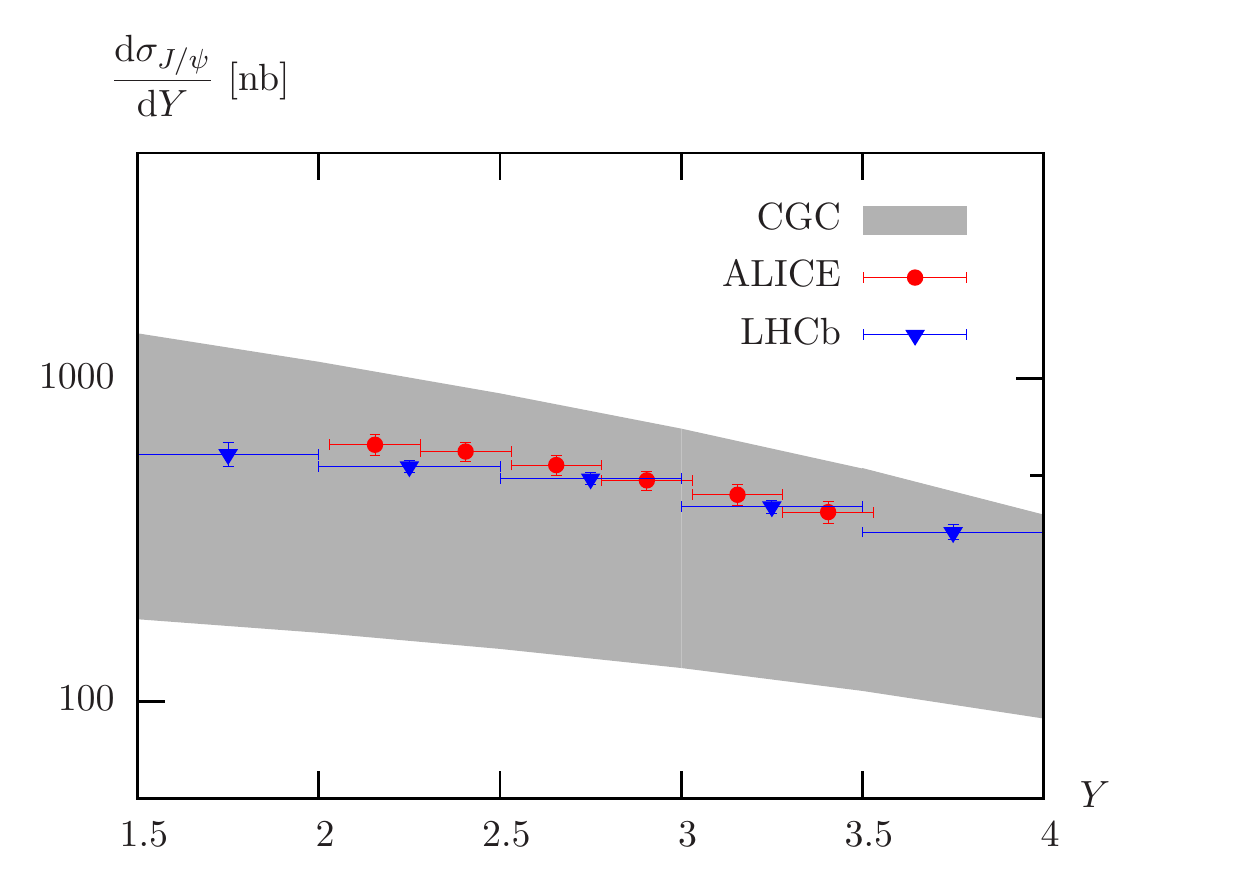}
	\caption{Differential $\Jpsi$ cross section as a function of $Y$ in pPb collisions. Data from Refs.~\cite{Abelev:2013yxa,Aaij:2013zxa}.}
	\label{fig:dsigma_dY_pPb_Jpsi}
\end{figure}

\begin{figure}[tb]
	\centering\includegraphics[width=\picwidth]{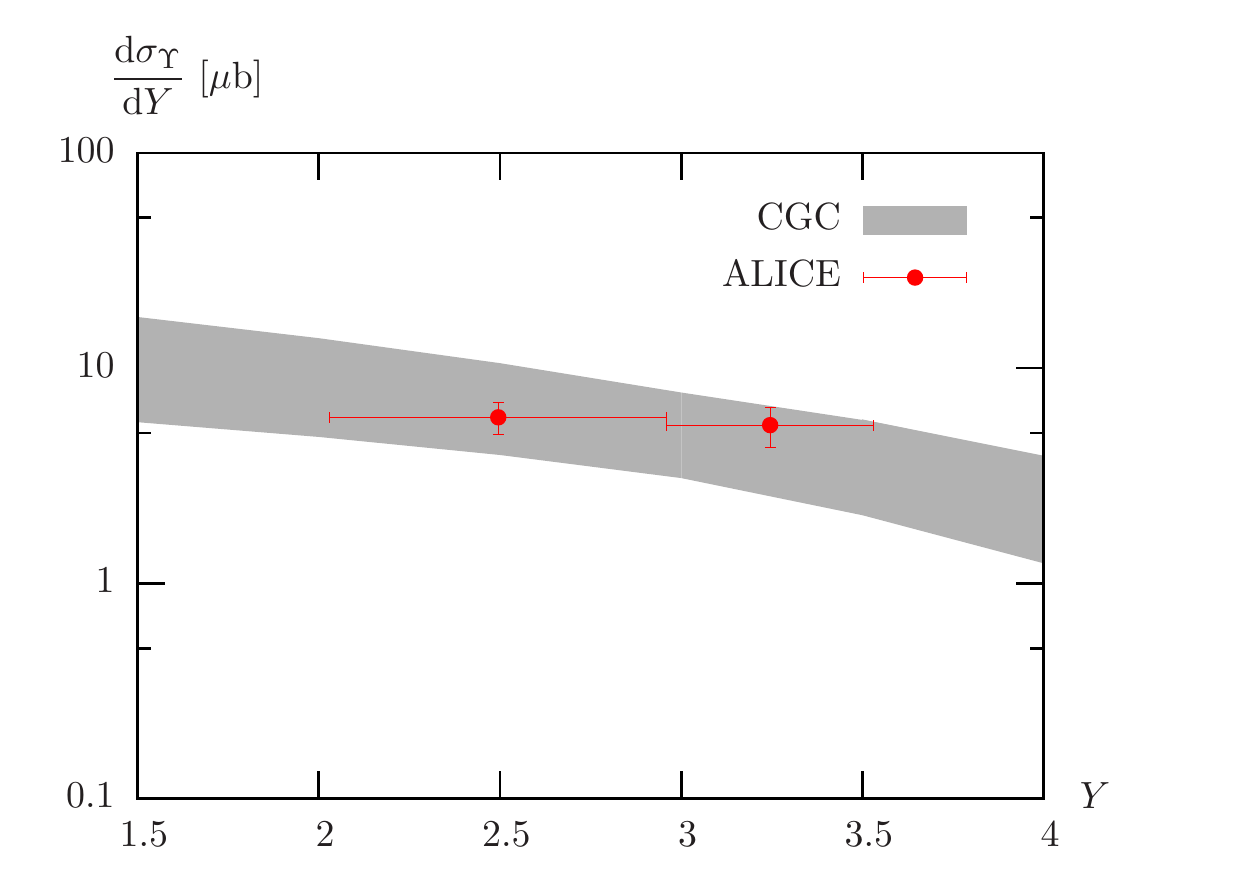}
	\caption{Differential $\Upsilon$ cross section as a function of $Y$ in pPb collisions. Data from Ref.~\cite{Abelev:2014oea}.}
	\label{fig:dsigma_dY_pPb_Upsilon}
\end{figure}

\begin{figure}[tb]
	\centering\includegraphics[width=\picwidth]{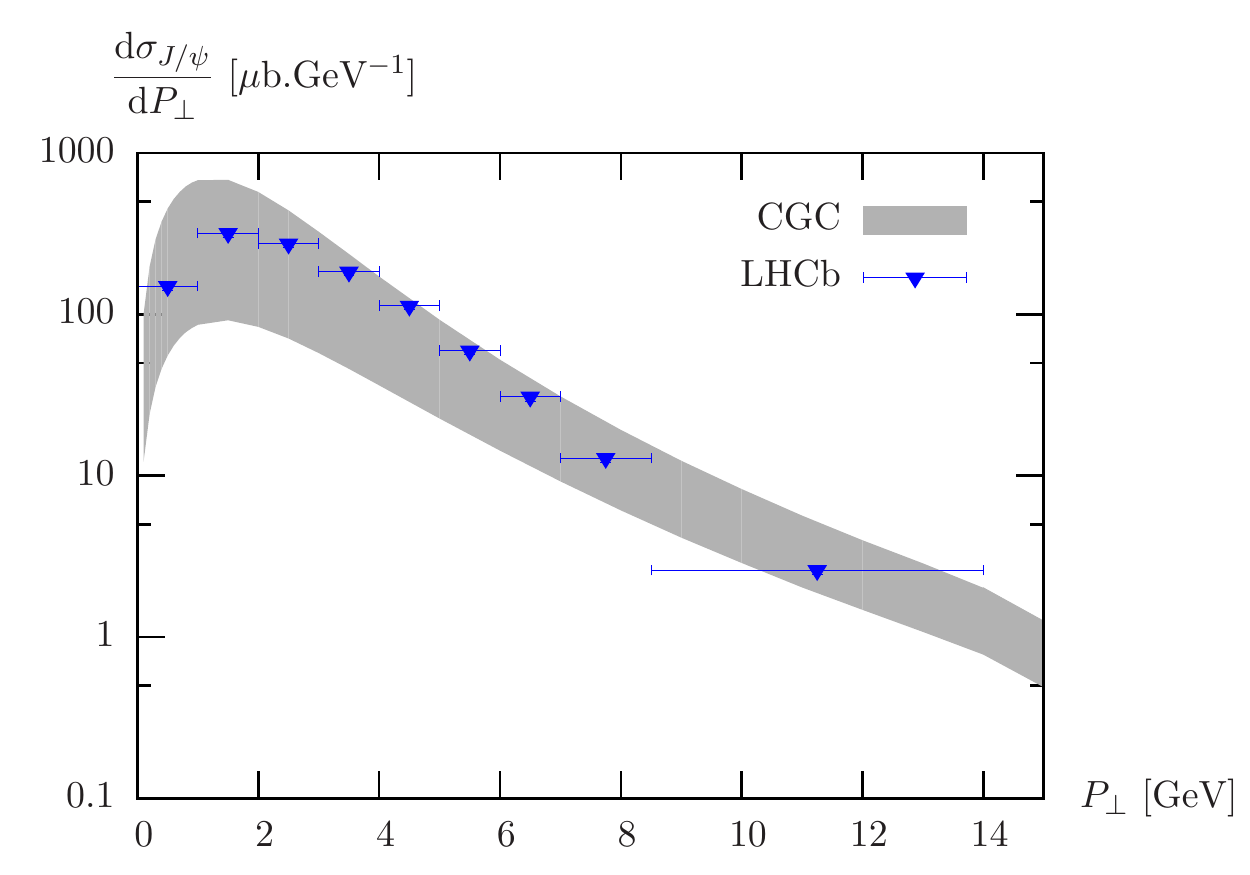}
	\caption{Differential $\Jpsi$ cross section as a function of $\ptt$ in pPb collisions ($1.5<Y<4$). Data from Ref.~\cite{Aaij:2013zxa}.}
	\label{fig:dsigma_dpT_pPb_Jpsi}
\end{figure}

\begin{figure}[tb]
	\centering\includegraphics[width=\picwidth]{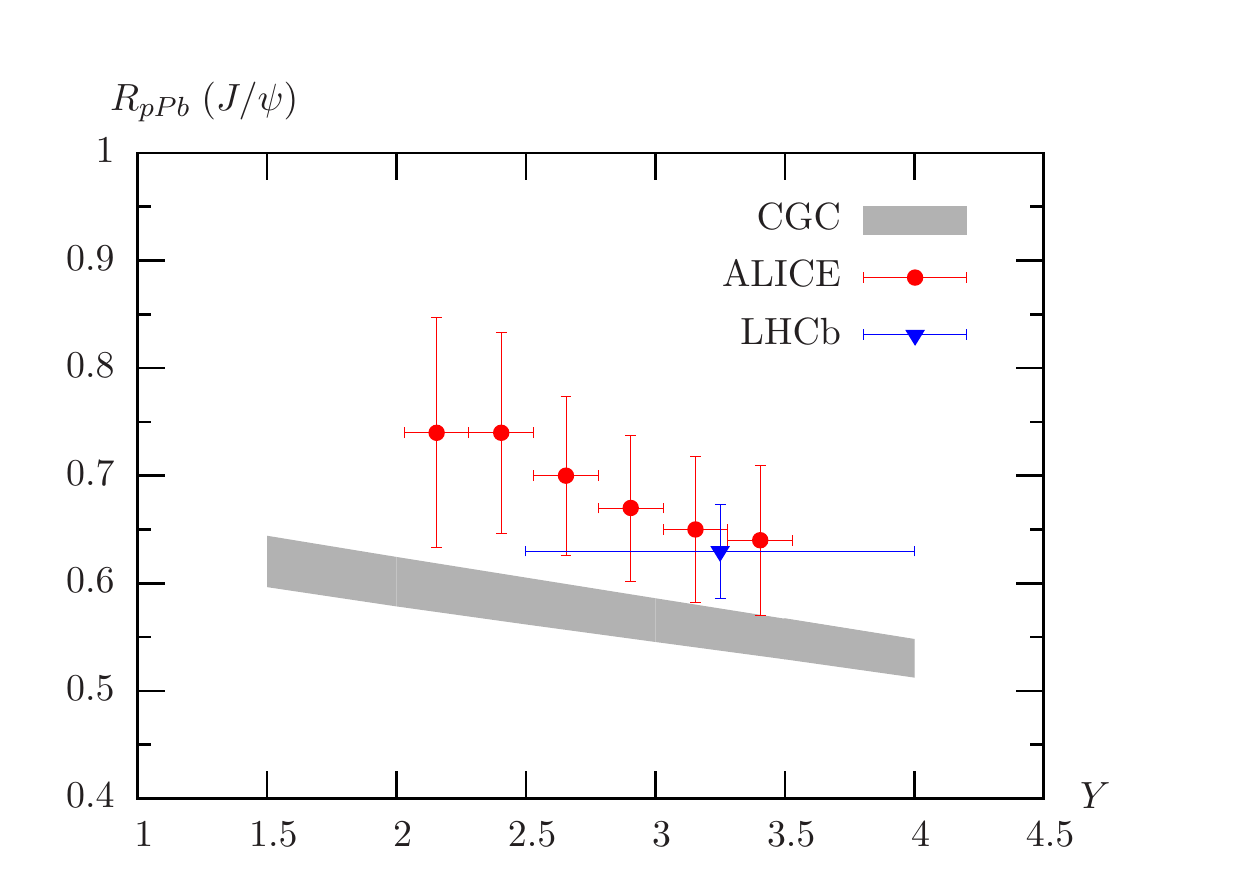}
	\caption{Nuclear modification factor for $\Jpsi$ production as a function of $Y$. Data from Refs.~\cite{Abelev:2013yxa,Aaij:2013zxa}}
	\label{fig:RpA_Y_Jpsi}
\end{figure}

\begin{figure}[tb]
	\centering\includegraphics[width=\picwidth]{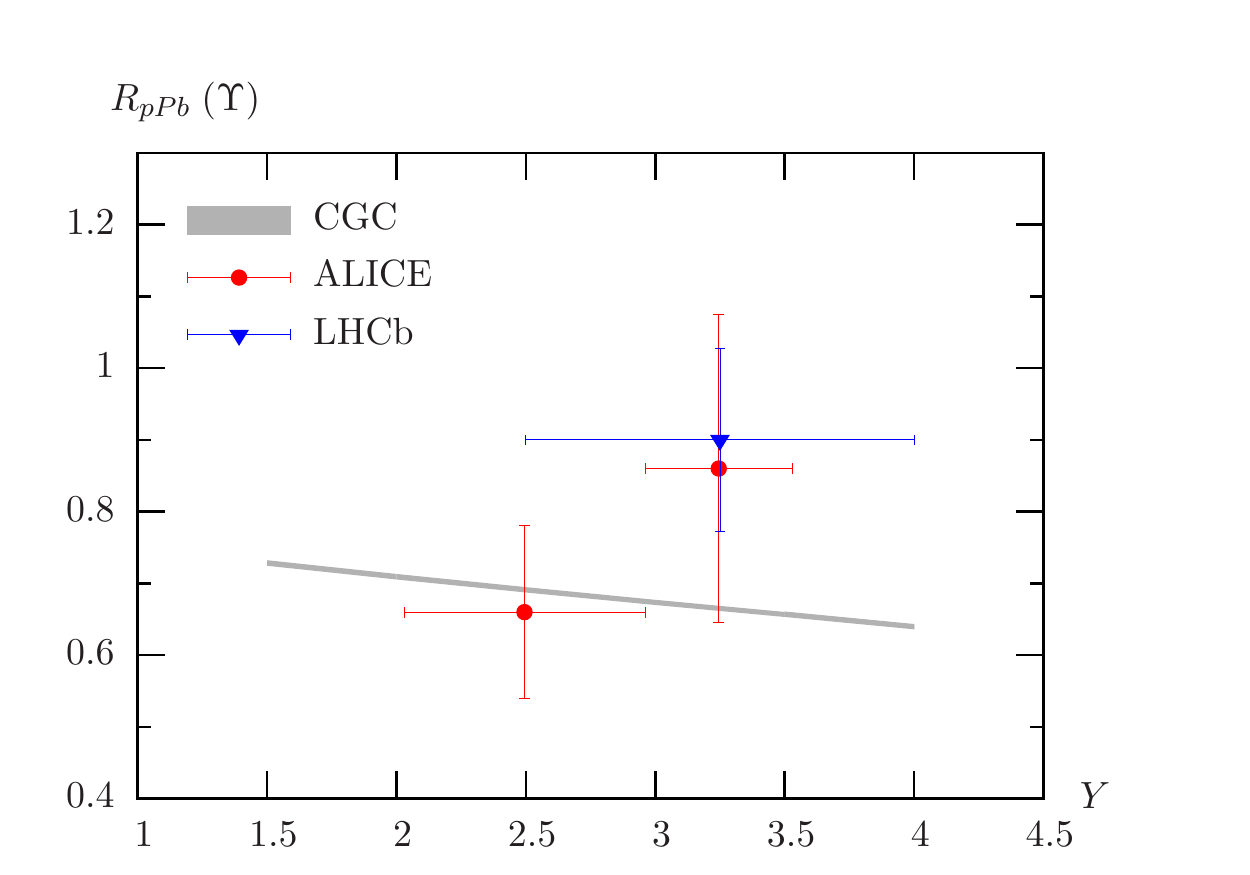}
	\caption{Nuclear modification factor for $\Upsilon$ production as a function of $Y$. Data from Refs.~\cite{Abelev:2014oea,Aaij:2014mza}.}
	\label{fig:RpA_Y_Upsilon}
\end{figure}
\begin{figure}[tb]
	\centering\includegraphics[width=\picwidth]{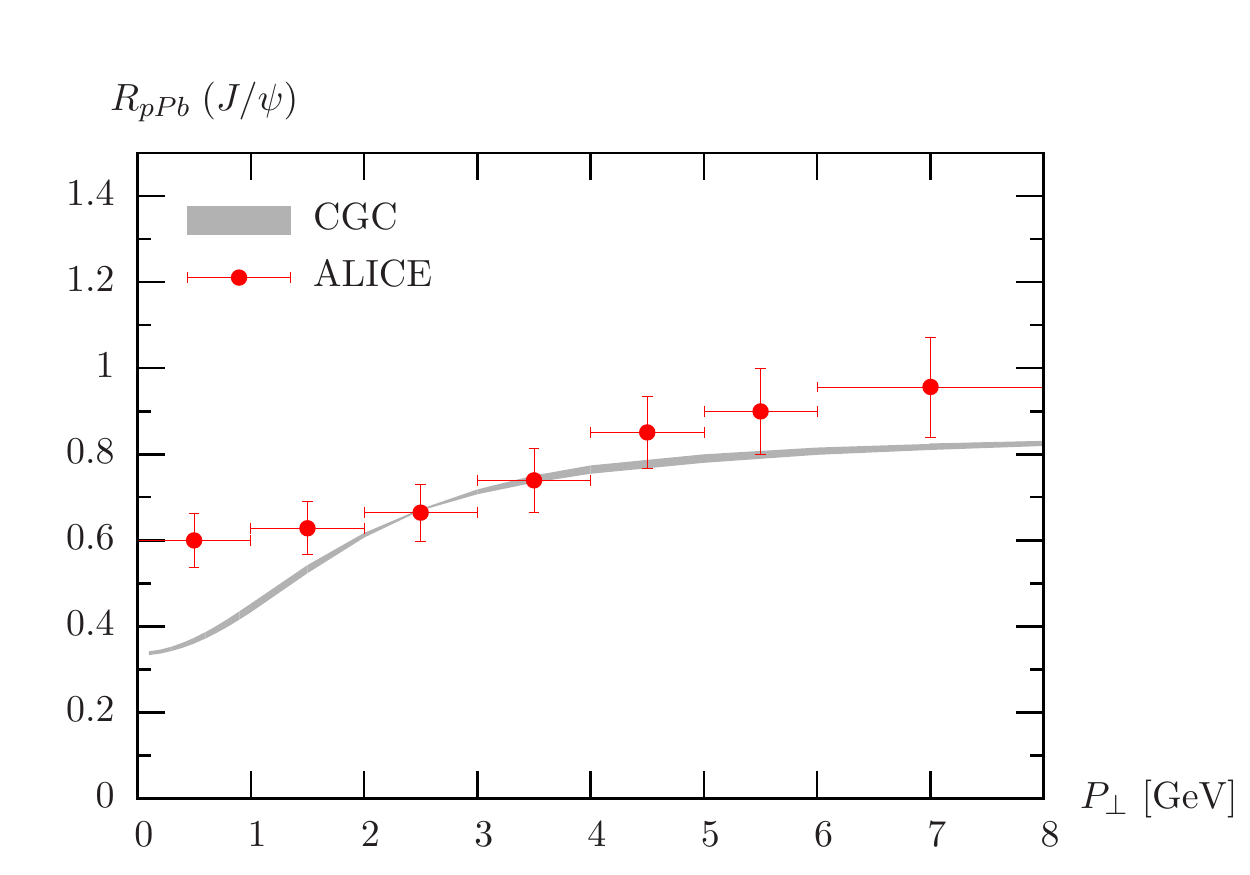}
	\caption{Nuclear modification factor for $\Jpsi$ production as a function of $\ptt$ ($2<Y<3.5$). Data from Ref.~\cite{DaCosta:2014tpa}.}
	\label{fig:RpA_pT_Jpsi}
\end{figure}

\begin{figure}[tb]
	\centering\includegraphics[width=\picwidth]{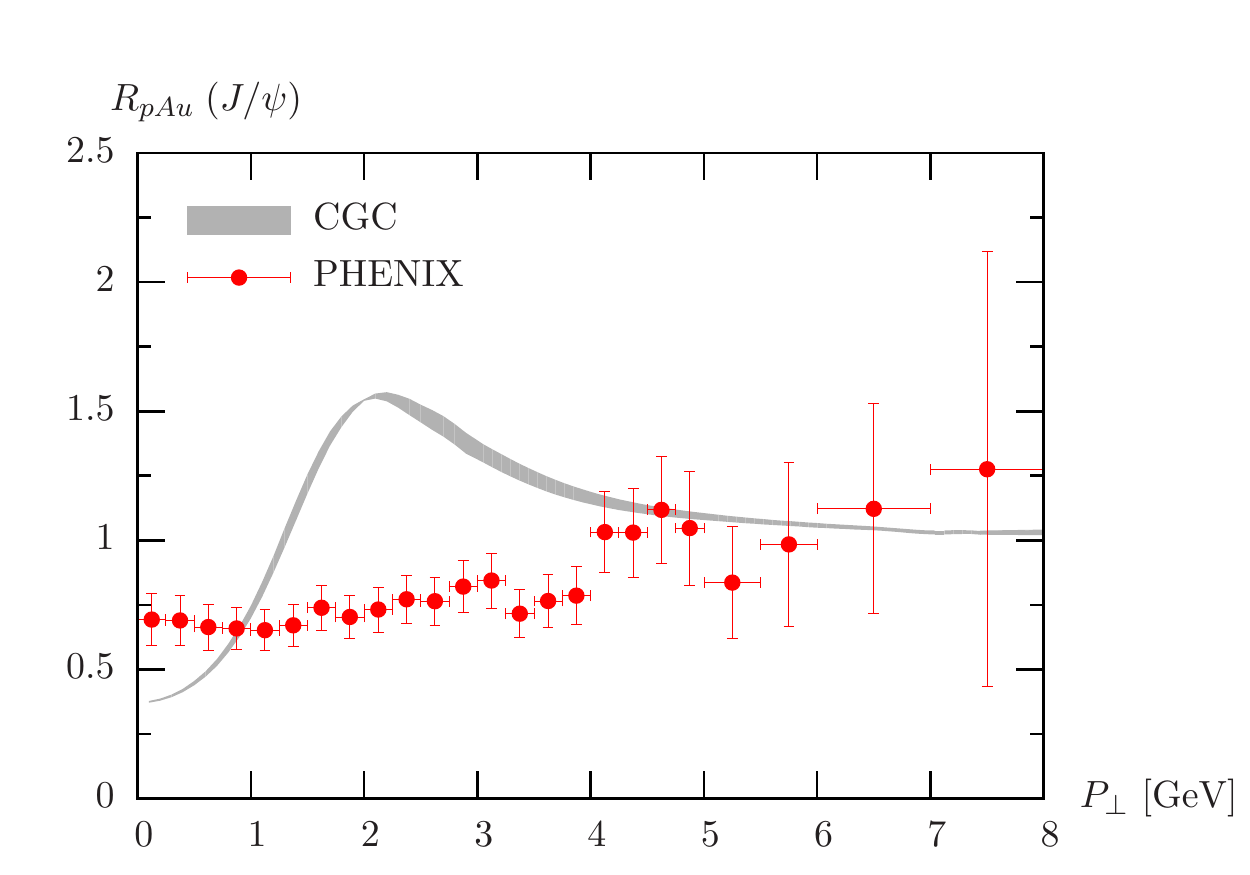}
	\caption{Nuclear modification factor for $\Jpsi$ production at RHIC energies as a function of $\ptt$ ($1.2<Y<2.2$). PHENIX data from Ref.~\cite{Adare:2012qf}}
	\label{fig:RpA_pT_Jpsi_RHIC}
\end{figure}

\subsection{Proton-lead collisions}

In \fig\ref{fig:dsigma_dY_pPb_Jpsi} we show the differential $\Jpsi$ cross section as a function of the rapidity in pA collisions at a center of mass energy of 5 TeV and compare with ALICE~\cite{Abelev:2013yxa} and LHCb~\cite{Aaij:2013zxa} data. As in the pp case, the slope of the data is compatible with the one predicted by our calculation but the normalization uncertainty is large.
In \fig\ref{fig:dsigma_dY_pPb_Upsilon} we show the same quantity for $\Upsilon$ meson production and compare our results with data from the ALICE Collaboration~\cite{Abelev:2014oea}. Here we can see that the normalization uncertainty of our calculation is again smaller than for $\Jpsi$ production, but there are only two data points to compare with, which limits the scope of this comparison.

In \fig\ref{fig:dsigma_dpT_pPb_Jpsi} we show the differential cross section as a function of the transverse momentum of the $\Jpsi$ in pA collisions at a center of mass energy of 5 TeV and compare with LHCb~\cite{Aaij:2013zxa} data. The agreement with data is good but it is not possible to see if our calculation overestimates the cross section at large $\ptt$ like it does for pp collisions since there is only one experimental bin between 8.5 and 14~GeV.

The normalization uncertainty from the charm quark mass and the factorization scale is 
quite large in the cross section itself. The modification of the production 
mechanism due to the stronger color fields in nuclei is, however, very little 
affected by these factors. Indeed, if the nuclear geometry is properly treated within the Glauber picture, the CGC framework provides a very robust prediction for the nuclear modification factor, defined as
\begin{align}\label{eq:defrpa}
R_{pA}= \frac{1}{A}\frac{\left . \ud \sigma/ \ud^2 
	\Pt \ud Y \right |_\text{pA}}
{\left . \ud\sigma/\ud^2 \Pt \ud Y \right |_\text{pp}} \; .
\end{align} 
This ratio has been extracted from the proton-proton and proton-nucleus 
cross sections by both the ALICE and LHCb collaborations. 
In \fig\ref{fig:RpA_Y_Jpsi} we show it in the case of $\Jpsi$ production as a function of the rapidity 
and compare with experimental data. We observe 
that our results are close to the 
data when taking into account all sources of uncertainty.
In particular, as already discussed in the Introduction, our calculation is 
much closer to the data than early CGC calculations in 
Ref.~\cite{Fujii:2013gxa}.
The nuclear modification factor for $\Upsilon$ production was also measured by the ALICE~\cite{Abelev:2014oea} and LHCb~\cite{Aaij:2014mza} collaborations, in two and one rapidity bins respectively. We show the comparison of these data with our results on \fig\ref{fig:RpA_Y_Upsilon}. Here no firm conclusions can be drawn at the moment since experimental uncertainties are quite large. 

Finally we look at the the behavior of $R_\text{pA}$ as a 
function of $\ptt$. 
Figure \ref{fig:RpA_pT_Jpsi} shows the behavior of $R_\text{pA}$ as a 
function of $\ptt$ for $\Jpsi$ production, compared to the measurement by the ALICE
collaboration~\cite{DaCosta:2014tpa}. Except at very 
low $\ptt$, the agreement of our calculation with this measurement 
is quite good, and the normalization uncertainty has almost completely canceled out. In Fig.~\ref{fig:RpA_pT_Jpsi_RHIC} we also compare our calculation to PHENIX
results~\cite{Adare:2012qf} at lower energies. While our calculation displays a quite strong ``Cronin'' peak that is not seen in the data, also here our 
results are closer to data than the previous CGC 
calculations~\cite{Fujii:2013gxa}.  The nuclear modification factor results at 
LHC energies, Figures~\ref{fig:RpA_Y_Jpsi} and \ref{fig:RpA_pT_Jpsi},
 are the main result of this
paper.

\section{Discussion and outlook}
\label{sec:disc}

At high enough transverse momentum, particle production in QCD processes 
should be well described by collinear factorization. Indeed many comparisons
of NLO calculations using nuclear parton distributions have been performed in the literature, see e.g. Ref.~\cite{Albacete:2013ei}. In general it seems that
the shadowing present in the standard EPS09~\cite{Eskola:2009uj} 
nuclear gluon distribution tends to underestimate
the nuclear suppression observed experimentally
in proton-nucleus collisions at lower transverse momenta. The collinear NLO calculation is in fact not that 
far from the physical picture of our work here. A leading order collinear 
calculation of heavy quark pair production results in a delta-function 
$\ptt$-distribution for the pair. To get a physically meaningful result
for the differential cross section, one must go to NLO, where an additional 
radiated gluon can carry the recoil transverse momentum. In a standard NLO
calculation, one integrates over the phase space of this gluon. At high
energy this additional phase space integration can contribute a large logarithm
$\sim \ln s$. The BK equation is, in this context, a resummation of
these logarithmically enhanced terms to all orders in $\as \ln s$. 
The large nuclear suppression in
the data could be taken as an indication that the nonlinear effects included in
BK indeed are significant for the $\Jpsi$.

In ``cold nuclear matter'' (CNM) energy loss calculations such as
Ref.~\cite{Arleo:2012rs}, the physical argument  behind the calculation
can also be described in a language
that is very similar to the one used here. The picture in the target rest frame 
is that of an incoming gluon splitting into a quark-antiquark pair. The pair
can then lose energy as it propagates through the target nucleus, leading to a 
nuclear suppression of the cross section. This picture does not rely on 
the eikonal approximation, making it possible to extend the formalism to 
higher $x_A$ and transverse momenta. The practical implementation of
the CNM model relies on a factorization of the quark pair production 
cross section and subsequent propagation through the nucleus. This 
factorization becomes questionable at very high energies in the target rest 
frame (very high $s$ and forward rapidities), when the typical target
transverse momentum scale $\qs$ is of the order of the charm quark mass. In this
limit the timescales of the quark pair formation and of the gluon radiation leading to the energy loss are not well separated. Indeed, in the CGC 
calculation the two are not factorized, but the pair production and propagation 
through the target nucleus are described coherently in the same formalism.

There is a normalization uncertainty in our calculation 
from the fact that we are using the dipole cross section~\cite{Lappi:2013zma} 
that was obtained for light flavors only. Requiring separately a good
description of the charm DIS cross section $F_2^c$ would require additional
parameters in the fit, e.g. as a separate
$\sigma_0$ for heavy quarks as in the 
AAMQS~\cite{Albacete:2010sy} parametrization.
This would improve our confidence in the normalization of the charm production 
cross section, but would be a significant work outside the scope of this paper.

We also used the very crude color evaporation hadronization model to 
obtain the $\Jpsi$ cross section from the 
quark pairs. Our estimate for the normalization uncertainty was rather conservative, as we varied the quark mass without adjusting the value of the color 
evaporation model constant $F_{\Jpsi}$ together with it.
The description of the transition from a quark pair to a $\Jpsi$ bound 
state  has seen  many recent theoretical advances in terms of non-relativistic QCD (NRQCD)~\cite{Bodwin:1994jh,Brambilla:2010cs}.
This has recently 
been successfully combined~\cite{Kang:2013hta,Ma:2014mri} with
the small-$x$ formulation for quark pair production.
It would be interesting to combine the newer proton and nucleus 
dipole cross sections used here with the NRQCD hadronization picture. This
is, however, left for future work.

Both of the uncertainties discussed above affect more seriously the 
separate proton-proton and proton-nucleus cross sections. They should mostly 
cancel in the nuclear modification factor $R_{pA}$. We leave a more 
detailed investigation of the separate cross sections
to future work. The main purpose of this
 paper has been to focus on the importance of a proper treatment of the 
nuclear geometry, including the use of a proton transverse area consistent
with the DIS fits used to obtain the dipole cross section. We have shown 
that when this is done using the dipole cross section parametrization
obtained in Ref.~\cite{Lappi:2013zma}, the resulting nuclear modification
ratio is much closer to the experimental data than in previous 
CGC estimates.

\section*{Acknowledgements}
B.~D. and T.~L. are supported by the Academy of Finland, projects 
267321 and 273464 and H.M by 
the Graduate School of Particle and Nuclear Physics.
This work was done using computing resources from
CSC -- IT Center for Science in Espoo, Finland.

\bibliography{../../bib/spires.bib}
\bibliographystyle{JHEP-2mod}

\end{document}